# Investigation of Resonance between HVDC-MMC Link and AC Network

I. Radecic, B. Filipovic-Grcic, P. Akiki, A. Xémard, B. Jurisic

*Abstract*-- HVDC networks offer several advantages over traditional HVAC systems, particularly for long-distance power transmission and integration of renewable energy sources, such as reduced losses and enhanced stability and control, but also increase the risk of oscillations. This study investigates electrical resonant phenomena associated with HVDC stations through numerical EMT simulations. The findings indicate that electrical resonance is primarily pronounced in weak networks with long cables, as confirmed by the Nyquist criterion applied to frequency responses. Two real cases were successfully simulated in the time domain by introducing network changes, such as temporary faults and alterations in network's power strength, to activate the identified resonances. Notably, in a strong network with short cables, electrical resonance occurred alongside interactions between the network and the converter's protection system. The analysis of voltage waveforms revealed that the amplitude of the induced resonant harmonic dissipates quickly, indicating sufficient damping in the network configuration. Furthermore, the study confirmed the network's sensitivity to changes in converter parameters modeled using available MMC model.

## I. Introduction

High Voltage Direct Current (HVDC) stations are now a well-established technology. However, their frequency-dependent modeling in numerical programs such as electromagnetic transient (EMT) programs remains an area of uncertainty. Specifically, understanding the frequency response of converters, particularly HVDC stations, is of great importance when evaluating the network from a stability perspective, where the issue of increased oscillations in networks rich in power electronic devices is frequently highlighted as one of the main causes of instability [1][2].

One of the fundamental steps in understanding oscillatory issues in networks dominated by power electronic devices is their classification. The most common classification is based on the frequency at which the system oscillates. Specifically, if the system oscillates at a frequency lower than the nominal (50/60 Hz), it can be categorized as subsynchronous, whereas if it oscillates at a higher frequency (above 50/60 Hz), it can be classified as supersynchronous. CIGRE TB 909 [1] offers a thorough method for classifying subsynchronous oscillations. However, there is no comparable classification available in existing documents for supersynchronous oscillations.

If electrical resonance is identified as a possible oscillatory concern, it is categorized as a subsynchronous issue according to [1]. Nonetheless, both practical experience and various other sources as [3] and [4] demonstrate that electrical resonance should be also classified as a supersynchronous phenomenon.

The analysis of electrical resonance relies on both analytical and numerical approaches. Analytical methods involve solving complex mathematical challenges, further complicated by the limited knowledge of converter control systems due to intellectual property protections [1]. Understanding these systems is essential for determining the frequency response, especially in voltage source converters (VSCs), where frequency behavior is heavily control dependent [1][2]. As a result, numerical tools are becoming more critical for modeling HVDC stations. Modeling solely at the fundamental frequency is inadequate for analyzing electrical resonance, therefore HVDC station model must account for frequency dependency across a wider spectrum.

Based on the issues regarding the classification of electrical resonance and the challenges in its analysis, an AC-DC network was modeled in EMTP program, where HVDC stations were modelled using the available EMTP model. The resonance phenomenon was analyzed on an HVDC-MMC link connected by cable to an AC network, with the study focusing on conditions that might lead to resonance. The modeled network was analyzed from the perspective of electrical resonance using the impedance-based method described in [1] and [5], which is based on the premise that the network can be divided into subsystems that may oscillate relative to each other. Thus, the presented analysis is grounded in obtaining the frequency response of each observed subsystem with respect to a selected point in the network. The frequency responses of individual parts of the network were obtained numerically by applying the small signal perturbation method described in [1] and [5], while the network's tendency toward electrical resonance was evaluated based on Nyquist's criterion.

The analysis explored the likelihood of resonance occurrences based on factors such as short-circuit capacity, AC cable length, and the parameters of the converter control loop. Through parametric and sensitivity analyses, the study aimed to identify critical scenarios where resonance could arise, highlighting configurations and operating conditions that may present risks to system stability.

The network topology and its parameters are presented in Section II. The procedure for obtaining the frequency response, as well as the analysis of the results, is outlined in Section III. Identified issues in the frequency domain were later successfully confirmed in the time domain by implementing changes in the network (faults, switching operations), those results are presented in Section IV. Finally, the conclusion is provided in Section V.

## II. CASE STUDY– NETWORK TOPOLOGY

For analysis of electrical resonance, a network model was developed in EMTP. Network topology is shown in Fig. 1. Two networks are represented by Thevenin equivalents, equivalent network 1 and 2, each with a single-phase and three-phase short circuit power (i.e. Short Circuit Capacity - SCC) of 10 GVA and a nominal voltage level of 400 kV. These networks are connected by a long single-core XLPE AC cable, followed by two Modular Multilevel Converter (MMC) stations interconnected with a DC cable, enabling the bidirectional transmission of 1 GW of nominal power before connecting to the second network.

This network topology was selected to trigger electrical resonance between the capacitive behavior of the AC cable and the inductive behavior of HVDC stations. Therefore, the length of the AC cable in the base case was selected to be 160 km to further emphasize the effect. However, due to the long AC cable, shunt compensation of 500 mH was implemented on each side. Additionally, all cables were modelled using EMTP wideband cable model.

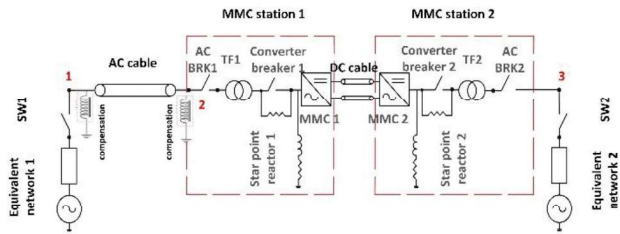

Fig. 1. Network topology

The MMC stations were modeled using an EMTP MMC model, with parameters set according to [6]. Key parameters for the MMC stations are shown in Table I. MMCs models involve a large number of IGBTs, complicating EMTP simulations. Detailed MMC models include thousands of IGBTs and require small time steps to accurately represent fast, simultaneous switching events. This heavy computational burden has led to the development of simpler Average Value Models (AVMs) that neglect switching details while still providing sufficient accuracy for dynamic simulations. AVMs significantly reduce computational resources and allow for larger integration time steps, resulting in faster computations. In EMTP program, four converter models of varying complexity are available, each offering a balance between accuracy and simulation speed. Model 1 (Full detailed) is the most complex and requires the longest computation time, while Model 4 (AVM) is the simplest. In this analysis, Model 3 was used, which is defined in [1] and [2] as sufficient for analyzing resonance phenomena. Model 3 represents an MMC arm using the averaged switching function concept of a half-bridge converter.

Notably, the converter transformer (TF1 in Fig. 1) was modeled non-linearly, incorporating its magnetization branch, whereas this feature was not included in the converter transformer (TF2 in Fig. 1) for MMC station 2. This was done mainly to speed up the process of reaching a steady state, which significantly reduces computation time. Including the magnetizing branch of both converter transformers, while contributing to the accuracy and detail of the network model, significantly slows down the analysis. The magnetization curve of TF1 is shown in Fig. 2.

TABLE I
EMT MMC MODEL PARAMETERS

| Parameter | Value |
|---|---|
| Type of model | Model 3 – switching function of arm |
| Configuration | Monopolar |
| Rated Power | 1000 MVA |
| AC primary/secondary voltage | 400/320 kV RMS LL |
| frequency | 50 Hz |
| DC pole-to-pole voltage | 640 kV |
| Transformer reactance | 0.18 p.u. |
| Transformer resistance | 0.001 p.u. |
| MMC arm inductance | 0.15 p.u. |
| Capacitor energy in each submodule | 40 kJ/MVA |
| Number of submodules per arm | 400 |
| Conduction losses of each IGBT/diode | 0.001 Ω |
| Star point reactor | 7700 Ω<br>6500 H |
| Inner current control time constant | 0.01 s |
| DC current maximum limit protection | 6 p.u. |

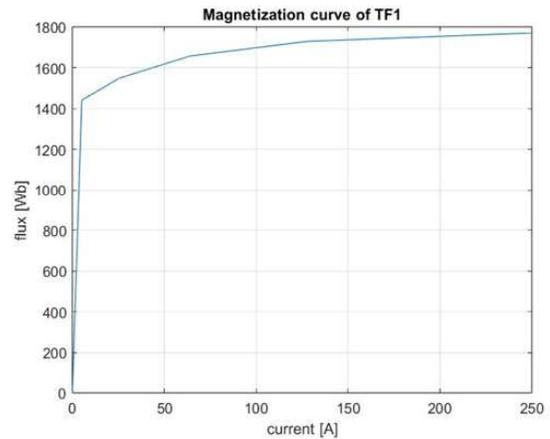

Fig. 2. Magnetization curve of TF1 in MMC Station 1

## III. FREQUENCY DOMAIN ANALYSIS

Once the network was created in EMT program, its frequency response can be obtained. In this paper, it was obtained by applying the small signal perturbation method described in [1] and [5]. The small signal perturbation method is based on the simple injection of a harmonic current at a selected point in the network during its steady state. This harmonic component will then be distributed according to the fundamental rule of current division, as illustrated in Fig. 3. The injected harmonic is thus divided between $I_1$ and $I_2$ and by additionally using the voltage $V$ at the point of perturbation injection, the impedance values of each network subsystem can be determined. This approach allows for the analysis of system dynamics by observing how the injected signal propagates through the network components, providing insights into the system's frequency response and stability characteristics.

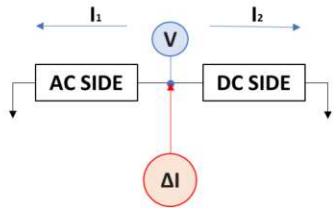

Fig. 3. Small signal perturbation method

The described procedure was therefore applied to the steady state values when transmitting power of 1 p.u. where the perturbation current is injected at two different locations, at Point 1 and Point 2, as shown in Fig. 1. Current harmonics were injected across a frequency range of 1 Hz up to 5000 Hz with varying steps. The impedance, calculated based on Ohm's law using the observed currents and voltages, was then interpolated to accelerate the process.

Observing from Point 1, impedance $Z_1$ contains only the Equivalent network 1, while impedance $Z_2$ includes everything else, such as the AC cable, two MMC stations, the DC cable, and the Equivalent network 2. Point 2 differs considering that the impedance of the AC cable is now included in impedance $Z_1$ and excluded from impedance $Z_2$, thus defining the case of interaction between the AC and the DC part of the network. In the following text, the term AC side will refer to impedance $Z_1$, and the term DC side will refer to impedance $Z_2$ despite the perturbation location to facilitate easier understanding.

It is important to note that the frequency analysis was conducted exclusively on phase A, and the presented results correspond to this phase alone. However, this should not pose an issue since the network is symmetrical. The obtained frequency responses for the base case (160 km AC cable and $SCC_1$=10 GVA) are displayed in Fig. 4 for Point 1, and in Fig. 5 for Point 2, respectively.

After obtaining the frequency responses for both locations, the results were validated by calculating the AC side response using frequency scan ($\Delta f$=1 Hz) in EMTP program described in [7]. This method does not apply to the DC side, as the system needs to be in a steady state for proper control operation. This is also one of the difficulties that occur during the frequency analysis of power electronics devices in numerical programs. Comparisons of the AC side frequency responses are shown in Fig. 6 and Fig. 7.

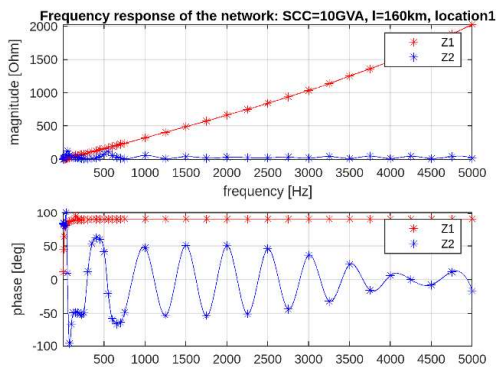

Fig. 4. Frequency response for the base case, Point 1

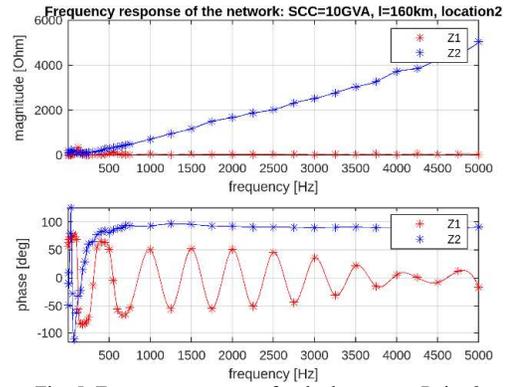

Fig. 5. Frequency response for the base case, Point 2

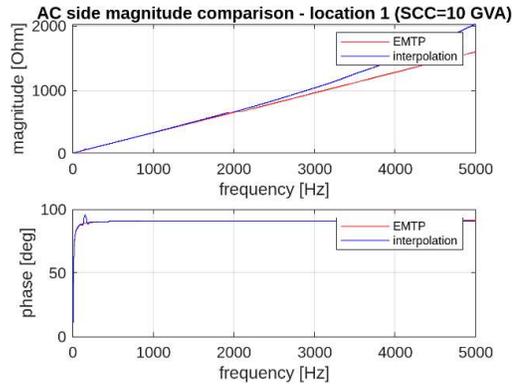

Fig. 6. Validation of perturbation method using frequency scan in EMTP program, Point 1

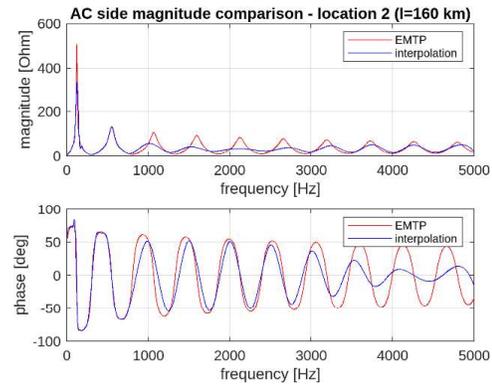

Fig. 7. Validation of perturbation method using frequency scan in EMTP program, Point 2

Reviewing Fig. 6 and Fig. 7, it was concluded that the frequency response of the AC side impedances obtained via the perturbation method closely matches the EMTP program frequency scan results, particularly for frequencies below 750 Hz, where interpolation is more accurate due to the smaller frequency step. Above 750 Hz, differences in magnitude arise due to larger frequency steps between each perturbation. A smaller time step would enhance accuracy but significantly increase simulation time. The authors consider this a minor detail, as frequency responses above 750 Hz are not used later in the analysis and no resonance conditions are present (as explained later in the text). Additionally, for Point 1, with just the Equivalent network 1 the response shows an inductance increasing with frequency, consistent with an R-L load model. More detailed frequency-dependent modeling of equivalent network could improve accuracy. For Point 2, which includes a 160 km AC cable, the response shows typical cable resonance behavior, with alternating inductive and capacitive modes at resonant frequencies.

Analogously to the validated AC sides, it can be considered that the DC sides are also well-obtained. This is further supported by the fact that the DC side at Point 2 exhibits inductive behavior, which aligns with the expected response based on [1], [2] and [8].

When the frequency responses are validated, the Nyquist criterion can be applied to assess the stability at specific frequency. There are generally two conditions to consider: first, there must be an intersection point where the impedance of AC side is equal to impedance of DC side ($\Delta Z=0$); second, the phase difference at the intersection frequency should exceed 180° ($\Delta \varphi \geq 180°$). It may be worth noting that the Nyquist criterion, while limited to linear systems and thus primarily applied in small-signal analysis, offers a valuable method for assessing stability. Despite its constraints, Nyquist's approach provides a clear stability criterion, making it highly useful in control systems analysis. Attempting to perform a similar stability analysis directly in the time domain would be considerably more complex and less straightforward, highlighting the Nyquist criterion's practical advantages.

If Nyquist's rule is applied to the frequency responses shown in Fig. 4 and Fig. 5, it can be concluded that at frequencies above 700 Hz, there is a significant difference in impedance amplitudes, making the occurrence of electrical resonance at these frequencies practically impossible (as first condition $\Delta Z=0$ cannot be satisfied). Therefore, further analyses are based on the network's frequency responses up to 700 Hz, which supports the accuracy of the results obtained through the implemented small signal perturbation method. Therefore, the frequency responses of the network up to 700 Hz are additionally presented in Fig. 8 and Fig. 9 for both locations.

Before applying the Nyquist criterion, a passivity rule suggests that unstable resonance is unlikely if the phase angle of impedances remains within -90° to 90° limit (nonpassive behavior of DC side is highlighted in red in Fig. 8 and Fig. 9). The non-passive behavior is more pronounced at Point 2, as it is closer to the MMC stations.

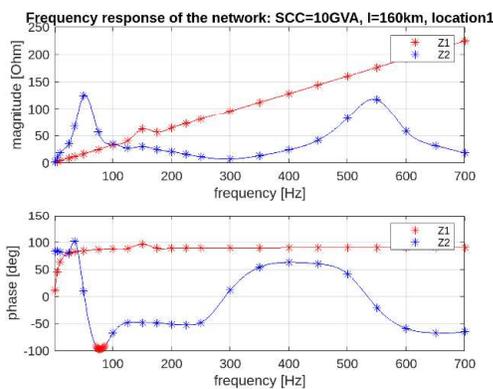
Fig. 8. Frequency response for the base case up to 700 Hz, Point 1

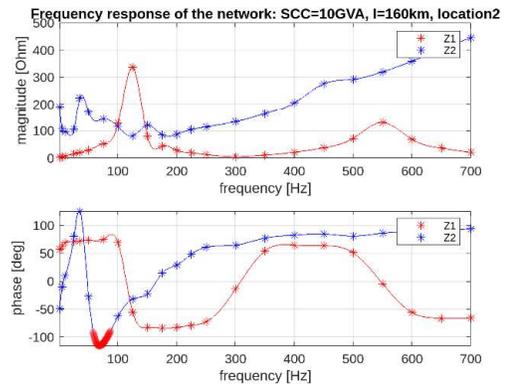
Fig. 9. Frequency response for the base case up to 700 Hz, Point 2

At Point 1, where non-passive behavior is observed, the application of the Nyquist criterion identifies a single magnitude intersection at approximately 90 Hz, where the phase difference surpasses 180°, indicating a potential for resonance. In contrast, despite the more pronounced non-passive behavior at Point 2, no frequencies meet the Nyquist criterion (as conditions $\Delta Z=0$ and $\Delta \varphi \geq 180°$ are not satisfied). Here, two intersection points of impedance magnitudes occur at around 100 and 150 Hz, but the phase difference does not exceed 180°, ruling out the possibility of resonance.

Since the network topology with base parameters left only a narrow margin for instability, a sensitivity analysis of the network's frequency response with altered parameters was conducted. It is additionally explained in the subsection of this section.

### A. Sensitivity Analysis

The length of the AC cable was adjusted to 40 km (compensated with 1500 mH) and 80 km (compensated with 1000 mH). Additionally, the network strength, specifically $SCC_1$, was reduced to 4.7 GVA for Equivalent network 1. The previously described frequency analysis and the procedure for obtaining it were repeated for each combination of network parameters.

Table II presents the outcomes derived from the application of Nyquist's criterion to the obtained frequency responses. An 'X' denotes instances where stability is maintained, while the frequency ranges in which instability may occur are indicated in other cases.

TABLE II
SENSITIVITY ANALYSIS

|  | Point 1 | Point 2 |
|---|---|---|
| AC cable length | $l$=160 km | |
| $SCC_1$=10 GVA | 86 – 91 Hz | X |
| $SCC_1$=4.7 GVA | 75 – 76 Hz | 76 Hz |
| AC cable length | $l$=80 km | |
| $SCC_1$=10 GVA | X | X |
| $SCC_1$=4.7 GVA | 81 – 84 Hz | X |
| AC cable length | $l$=40 km | |
| $SCC_1$=10 GVA | X | X |
| $SCC_1$=4.7 GVA | X | X |

According to Table II, electrical resonance is more prevalent in weak networks with long cables compared to strong networks with shorter cables, this is also reported in [9]. Even though shorter cables shift the resonant frequency higher, thus creating more magnitude intersection points, the phase difference remains inadequate to meet the Nyquist criterion. It is essential to note that these observations are specific to the examined network topology, so caution is warranted in generalizing the findings.

The gained insights have improved the understanding of network behavior across the 1 Hz to 5000 Hz frequency spectrum and facilitated resonance induction in the time domain. Examples of the network's behavior during resonance in various scenarios are discussed in Section IV., which shifts focus to analyzing temporary overvoltages (TOVs) resulting from triggered resonance in the time domain.

## IV. Time Domain Analysis

The goal of this section is to analyze the TOVs in the time domain that are caused by electrical resonance. Given that sensitivity analysis has identified a tendency toward resonance in the network topology with a long AC cable ($l$=160 km) and a weak network ($SCC_1$=4.7 GVA), this topology was selected as prone to electrical resonance. Two cases of electrical resonance were successfully triggered in such networks: one by applying a single-phase to ground fault and the other by changing the network's $SCC$. Based on this, it could be concluded that electrical resonance primarily occurs in weak networks with long cables. A unique case was observed where electrical resonance and oscillations in the converter's protection system occurred simultaneously, interacting with the electrical system. This, however, took place in a completely different setting, defined by a strong network and a short cable.

The subsections of this section provide separate analyses for each network type: weak networks with long cables and strong networks with short cables.

### A. Weak Network with Long Cables

A weak network with a long cable is characterized by an AC cable length of 160 km and an $SCC$ of 4.7 GVA for Equivalent network 1. Previously, a sensitivity analysis identified the possibility of instability at frequencies around 75-76 Hz. The observed instabilities need to be triggered in some way. Specifically, by simulating a temporary single-phase fault to ground at Point 1 lasting 500 ms, voltage oscillations were observed at both ends of the AC cable (Point 1 and Point 2) after the fault clearance, with stronger oscillations noted closer to the converter (Point 2). The observed oscillations at Point 2 are shown in Fig. 10.

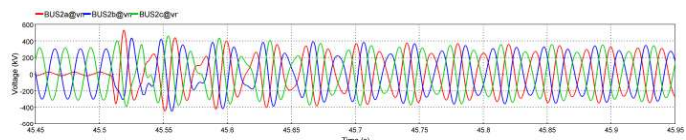
Fig. 10. Voltage oscillations at Point 2 after the one-phase fault clearance in weak network with long AC cable

To confirm that the oscillations were caused by electrical resonance, a Fast Fourier Transform (FFT) analysis of the phase A voltage waveform at Point 2 was performed. The results of the FFT analysis are shown in Fig. 11. Based on the results, as expected, the fundamental harmonic of 50 Hz with a dominant amplitude was observed. Additionally, a 76 Hz harmonic with an amplitude of 12.78 kV was detected. This observed frequency of 76 Hz corresponds to the previously identified unstable frequency range shown in Table II, confirming that electrical resonance occurred in the weak network with a long AC cable after the single-phase fault. The induced harmonic adds to the fundamental harmonic, causing TOV. However, the configured system possesses sufficient damping, allowing the induced harmonic to dissipate rapidly within just a few seconds.

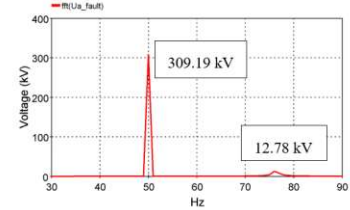
Fig. 11. FFT analysis of phase A voltage waveform at Point 2 after the single-phase fault clearance in weak network with long AC cable

The described oscillation was also successfully triggered by simulating a change in $SCC_1$ from 10 GVA to 4.7 GVA, resulting in voltage oscillations after reaching the new steady state. The frequency response of the network illustrated in Fig. 12 can explain this phenomenon. In the case of a strong network, the Nyquist criterion is not satisfied. However, when $SCC_1$ is reduced to 4.7 GVA, an unstable region emerges, indicating a change in the frequency response of the AC side. However, the resulting oscillations also dissipate very quickly.

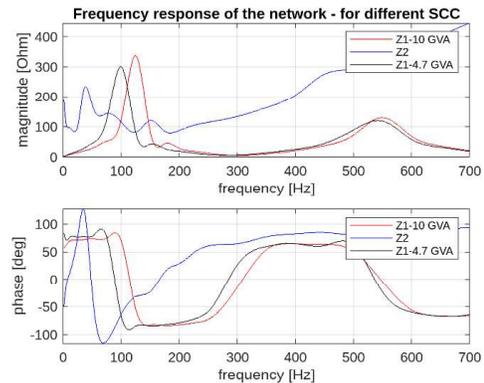
Fig. 12. Change in the frequency response due to the change of $SCC_1$

The evidence of oscillations suggests that the specified frequency range of around 76 Hz, while satisfying the Nyquist criterion, is not in an unstable region, but rather on the edge of instability. The primary cause of the error in the frequency response is attributed to interpolation. This situation indicates that although the system meets stability criteria, it may still face some instability under certain conditions, especially due to its sensitivity to parameter changes.

### B. Strong Network with Short Cables

Previously, a sensitivity analysis did not identify the potential for instability with shorter cables, leading to the conclusion that electrical resonance is not possible in strong networks with short cables. However, this time domain simulations successfully confirmed oscillations in a network with an AC cable length of 10 km (uncompensated) and an $SCC_1$ of 10 GVA.

A single-phase to ground fault was simulated, lasting for 500 ms, which resulted in oscillations within the system. The oscillation of the phase voltages during the fault at Point 2 is illustrated in Fig. 13, while after the fault clearance in Fig. 14.

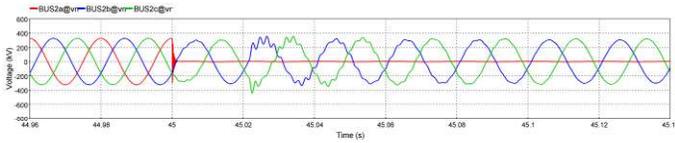

Fig. 13. Recorded oscillation of the voltage at Point 2, during the fault

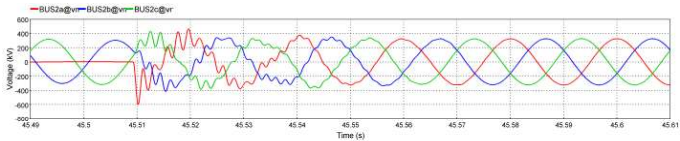

Fig. 14. Recorded oscillation of the voltage at Point 2, after the fault clearance

The waveforms presented are particularly interesting because, unlike the previous cases in weak networks with long cables, where oscillations only occurred after a fault and not during it, this scenario shows a different behavior. An FFT analysis was again performed on the displayed waveforms, with the results shown in Fig. 15 during the fault and after the fault, respectively.

It was observed that a single harmonic was not identified by the FFT analysis, which is characteristic of electrical resonance, as noted by [3]. This lack of a dominant harmonic suggests that the system may exhibit complex interactions among multiple frequencies. To explain the emergence of certain harmonics, a frequency analysis of the observed network topology was conducted. This analysis is crucial for understanding the behavior of the system and diagnosing potential resonance issues.

The results of the frequency analysis at Point 2 are shown in Fig. 16. The obtained frequency response was further analyzed using the Nyquist stability criterion with the aim of explaining the emerging harmonics.

By examining the frequency response of the network, it is noted that electrical resonance may occur at approximately 600 Hz as the Nyquist criterion is satisfied, which can explain the previously observed harmonic at 600 Hz after clearance of the fault (Fig. 15b). What was recorded was the occurrence of other harmonics during and after the fault, which cannot be explained by the presence of electrical resonance. Specifically, these harmonics appear 20 ms after the fault, corresponding to the time required for the converter protection systems (implemented in the MMC model as described in [6]) to activate. Therefore, the fault was repeated with the protective systems disabled. The voltage waveforms for the same fault and network configuration are presented in Fig. 17 and 18, showing the behavior during and after the clearance of the fault.

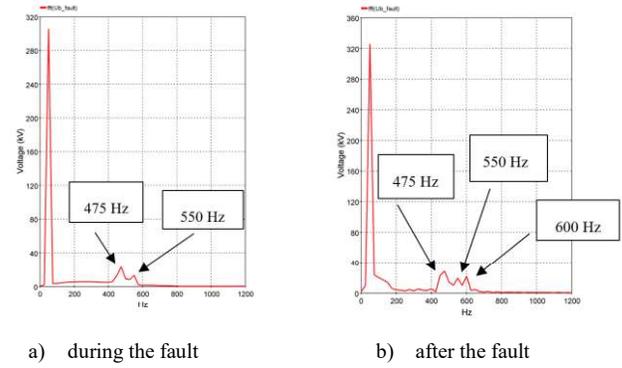

a) during the fault    b) after the fault

Fig. 15. FFT analysis of phase A voltage waveform at Point 2 in strong network with short AC cable

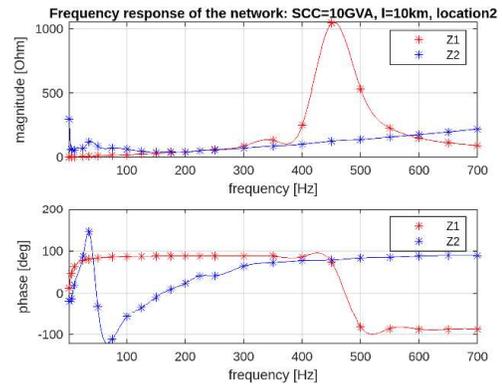

Fig. 16. Frequency response of the network characterized as strong with short AC cable, Point 2

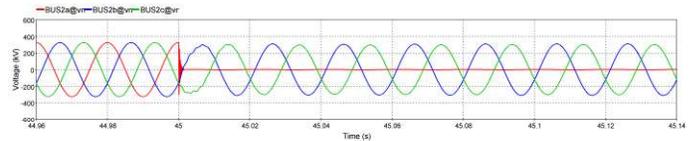

Fig. 17. Recorded oscillation of the voltage at Point 2, during the fault; protection system disabled

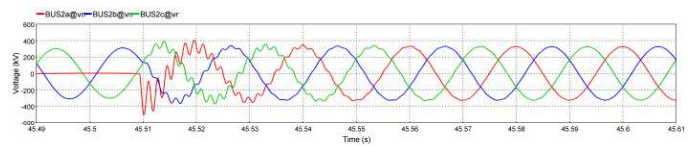

Fig. 18. Recorded oscillation of the voltage at Point 2, after the fault; protection system disabled

By comparing Fig. 17 and 18 with Fig. 13 and 14, it was observed that the resulting oscillations were reduced, completely disappearing during the fault. This observation was confirmed by the FFT analysis of the voltage waveforms. The comparison of voltage waveforms and FFT analysis reveals that the harmonics at 475 and 550 Hz were induced by the interaction between the network and the converter's protection system, as they were absent when the protection system was disabled (Fig. 19). Additionally, 600 Hz electrical resonance was confirmed using the Nyquist criterion, with the harmonic magnitude varying based on the status of the protection system: 22 kV with the system enabled and 12 kV without it. This indicates that the protection system has not only caused additional harmonics but also amplified those associated with electrical resonance.

This example shows that two oscillation problems can occur simultaneously, complicating the oscillation analysis and emphasizing the significant influence of converter parameters. To investigate this, a single-phase to ground fault was simulated again with a slower inner current control time constant of both MMCs (increased from 0.01 s to 0.02 s). The voltage waveform at Point 2 after the fault clearance with the protection system enabled is illustrated in Fig. 18. When comparing this waveform to waveform shown in Fig. 14, it is evident that the oscillations at Point 2 have a longer duration. This suggests that the slower control loop adversely affects the duration of these oscillations.

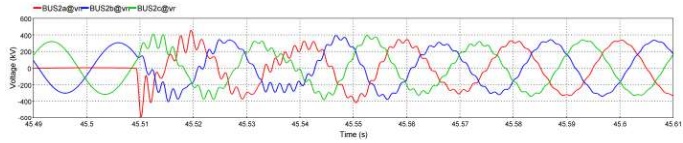

Fig. 18. Recorded oscillation of the voltage at Point 2, after the fault clearance with slower inner current control time constant

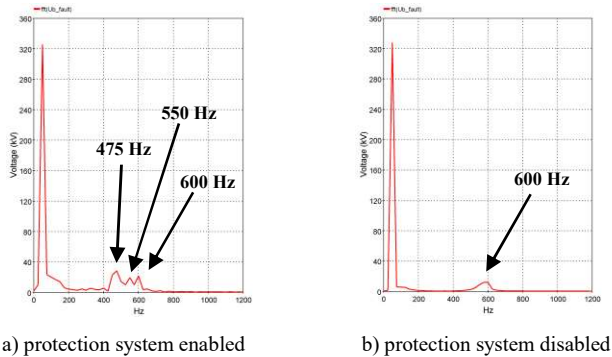

a) protection system enabled     b) protection system disabled

Fig. 19. Frequency response of the network from the Point 2, $S_{CC1}$=10 GVA, cable length $l$=10 km

Recorded sensitivity of the system to such small, but significant changes in converter parameters emphasizes the issue of modelling HVDC systems using models without considering the true parameters of the converter due to the manufacturer's IP protection.

## V. CONCLUSIONS

Studying mixed AC-DC networks is complex, but numerical tools like EMT programs facilitate the analysis of control-loop interactions, system nonlinearity, and resonant phenomena. The small signal perturbation method for obtaining frequency responses is intricate and time-consuming, often affected by interpolation methods, yet it provides crucial insights into network behavior across frequencies.

Initial assumptions suggested electrical resonance might occur between capacitive networks and inductive converters. However, significant impedance differences render this unlikely according to the Nyquist criterion. Resonance is primarily confined to a narrow frequency range at lower frequencies, where the network behaves inductively, and the converter exhibits capacitively non-passive characteristics. Sensitivity analysis revealed that weak networks with long cables are more susceptible to resonance, particularly highlighting non-passive behavior in the DC portion of the network.

The investigated TOVs are important for the system as they may stress equipment insulation and surge arresters. If excessive, they can degrade insulation or overstress arresters, reducing their effectiveness, or they can cause saturation problems and heating effects. Time domain simulations showed that while high TOVs are initially induced, they quickly dampen in such network configurations. However, dampening depends on many parameters so conclusions cannot be generalized. An interesting case emerged in a strong network with a short cable, where simultaneous oscillations were observed due to electrical resonance and interactions with the protection system, which increased overvoltages and underscored the network's sensitivity to minor changes.

This paper not only confirms the necessity of frequency-dependent network modeling with a focus on electrical components, but also highlights the critical role of accurately modeling control loops, as they can significantly impact on the network's oscillatory behavior.